\documentclass[a4paper,fleqn]{cas-dc}

\usepackage{graphicx,subcaption}
\usepackage{rotating}
\usepackage{lscape}
\usepackage{multicol}
\usepackage[justification=centering]{caption}
\usepackage{layout}
\usepackage{caption}
\usepackage[normalem]{ulem}
\usepackage[utf8]{inputenc}
\usepackage{aas_macros}
\usepackage{threeparttable}
\usepackage[authoryear]{natbib}
\usepackage{epstopdf}
\newcommand{\gm}{$\gamma$}
\def\tsc#1{\csdef{#1}{\textsc{\lowercase{#1}}\xspace}}
\tsc{WGM}
\tsc{QE}

\begin{document}
\UseRawInputEncoding
\let\WriteBookmarks\relax
\def\floatpagepagefraction{1}
\def\textpagefraction{.001}

\shorttitle{Multiwavelength Variability of OP\,313}    

\shortauthors{N.M.PN}  

\title [mode = title]{Deciphering the Multi-Wavelength Flares of the Most Distant Very High-Energy (\(>\)100 GeV) $\gamma$-ray Emitting Blazar}  



%

\author[1,2]{P N Naseef Mohammed}[orcid=0000-0003-0545-5998]

\cormark[1]


\ead{naseefmhd@farookcollege.ac.in}



\affiliation[1]{organization={Farook College},
            city={Calicut},
            postcode={673632}, 
            state={Kerala},
            country={India}} 
\affiliation[2]{organization={University of Calicut},
            city={Malappuram},
            postcode={673635}, 
            state={Kerala},
            country={India}}

\author[1,2]{T. Aminabi}[]
\ead{thekkothaminabi@gmail.com}

\author[2]{C. Baheeja}[]

\ead{baheeja314@gmail.com}

\author[3,4]{S Sahayanathan}[]
\ead{sunder@barc.gov.in}

\author[5]{Vaidehi S. Paliya}[]
\ead{vaidehi.s.paliya@gmail.com}

\author[2]{C D Ravikumar}[]
\ead{cdr@uoc.ac.in}


\affiliation[3]{organization={Bhaba Atomic Research Centre},
            city={Mumbai},
            postcode={400085}, 
            state={Maharashtra},
            country={India}}
\affiliation[4]{organization={Homi Bhabha National Institute},
            city={Mumbai},
            postcode={400094}, 
            state={Maharashtra},
            country={India}}
\affiliation[5]{organization={Inter-University Centre for Astronomy and Astrophysics (IUCAA),SPPU Campus},
            city={Pune},
            postcode={411007}, 
            state={Maharashtra},
            country={India}}

\cortext[cor1]{Corresponding author}


\nonumnote{}

\begin{abstract}
This study analyzes the multi-wavelength flaring activity of the distant flat spectrum radio quasar (FSRQ) OP 313 (z=0.997) during November 2023 to March 2024, using data from Fermi-Large Area Telescope, Swift X-ray Telescope, and Ultraviolet and Optical Telescope. The analysis highlights two significant {very high energy(VHE)} detection epochs and GeV gamma-ray flaring episodes, providing insight into jet emission processes and radiative mechanisms. Key findings include broadband spectral energy distribution (SED) evolution, including enigmatic X-ray spectral changes. Modeling of the multi-wavelength SED with a one-zone leptonic radiative processes attributes the emissions to synchrotron radiation, Synchrotron Self-Compton (SSC), and External Compton (EC) mechanisms, with torus photons as the primary source for EC processes. The results suggest that the gamma-ray emitting region lies outside the broad-line region but within the dusty torus. Furthermore, we find that the radiated power is significantly smaller than the total jet power, suggesting that most of the bulk energy remains within the jet even after passing through the blazar emission zone. These findings advance our understanding of particle acceleration, jet dynamics, and photon field interactions in FSRQs.

\end{abstract}

\maketitle
\section{\label{sec:intro}Introduction}
Blazars, a subset of Active Galactic Nuclei (AGN) and are one of the most energetic sources in the universe. They are characterized by rapid variability, high radio-to-X-ray polarization, and relativistically beamed jet emission \citep[e.g.,][]{1995PASP..107..803U}. These features make blazars interesting objects for astrophysical research, and their study can provide valuable insights into the mechanisms of jet production, particle acceleration, and the environments of supermassive black holes \citep[see, e.g.,][for a recent review]{2019ARA&A..57..467B}.


The optical spectrum of blazars often shows broad emission lines or a featureless continuum. Accordingly, they are classified into Flat Spectrum Radio Quasars (FSRQs) and BL Lacertae objects with the later associated with the ones with no/weak emission lines \citep{2012ApJ...748...49S,2013ApJ...764..135S}.Besides this, the spectral energy distribution (SED) of blazars spans across a wide range of wavelengths, from radio to very high-energy (VHE) gamma rays and are predominantly non-thermal in nature. It exhibits a double-peaked structure with the low-energy component generally attributed to synchrotron radiation from 
relativistic electrons in the jet, while the high-energy component is generally modeled as the inverse Compton scattering of low energy photons. The plausible target photons for the inverse Compton process can be the synchrotron photons themselves (synchrotron self-Compton SSC) or the photons external to the jet(external Compton EC) \citep{1998MNRAS.301..451G,1974ApJ...192..261J,1992ApJ...397L...5M,2000ApJ...545..107B,2017MNRAS.470.3283S}. 
Furthermore, based on the location of the synchrotron peak, blazars have also been categorized as low-synchrotron peaked (LSP, synchrotron peak frequency $\nu^{\rm pk}_{\rm syn}<10^{14}$ Hz), intermediate synchrotron peaked \mbox{($10^{14}<\nu^{\rm pk}_{\rm syn}<10^{15}$ Hz)}, and high-synchrotron peaked ($\nu^{\rm pk}_{\rm syn}>10^{15}$ Hz) blazars \citep[][]{2010ApJ...716...30A}.

FSRQs are mostly LSP-type sources, and the high-energy emission peaks at the MeV$-$GeV band. Accordingly, they exhibit a soft \gm~ray spectrum \citep[cf.][]{2020ApJ...892..105A}. This is also aligned with the fact that only a handful of FSRQs have been detected at very-high energies \citep[VHE, $>$100 GeV, cf.][]{2008Sci...320.1752M,2015ApJ...815L..22A,2020A&A...633A.162H}. A VHE detection of an FSRQ also provides clues about the location of the \gm-ray emission region, since the intense broad line region (BLR) photon field that can serve as a reservoir of the seed photon for the inverse Compton scattering can also absorb the \gm-ray photons via pair production process. Hence, the \gm-ray spectrum will fall sharply at VHE and the presence of a relatively hard spectrum indicates the emission region is beyond the BLR photon field \citep[e.g.,][]{2003APh....18..377D,2016ApJ...821..102B}. Apart from this, the VHE photons also undergo absorption due to pair production process with the extragalactic background light (EBL). As the distant sources encounter a relatively long column
length of EBL compared to the nearby ones, VHE detection of distant ($z>0.5$) FSRQs is challenging \citep{2013ApJ...763..145D,2022ApJ...941...33F,2020NatAs...4..124B}.


Understanding the causes of the dramatic flux variations observed from blazars has become an active area of research in the last two decade \citep[e.g.,][]{2010Natur.463..919A,2012ApJ...760...69N,2013ApJ...773..147J,2015ApJ...803..112P,2015ApJ...803...15P,2018A&A...619A.159M,2021MNRAS.504..416S}. The good photon statistics obtained during the elevated activity phase can provide us clues about the dynamic nature of these objects. Combining with the fact that FSRQs have been detected at VHE primarily during flaring episodes, an exhaustive multi-wavelength study of such periods are crucial to address outstanding questions as briefly elaborated above.

Fermi Large Area Telescope (LAT) observed an episode of GeV flaring activity from the distant FSRQ OP 313 \mbox{($z=0.997$)} in 2023 November \citep[][]{2023ATel16356....1B}. The followup observations with the Large-Sized Telescope (LST-1, a prototype of the Large-Sized Telescope for the Cherenkov Telescope Array Observatory) revealed a significant VHE detection during 2023 December 11$-$14 \citep[][]{2023ATel16381....1C}. Another \gm-ray flaring episode of even higher amplitude and harder \gm-ray spectrum was observed during 2024 February$-$March \citep[][]{2024ATel16497....1B}. This \gm-ray flaring activity of OP 313 presented a unique opportunity not only to explore the VHE emitting jet environment but also to improve our understanding of the EBL attenuation.

This paper presents a comprehensive multi-wavelength analysis of OP 313, focusing on the time period from 2023 November 12 to 2024 March 31. We utilized all publicly available multi-wavelength data taken with the Fermi-LAT, Swift X-ray Telescope (XRT), and Swift Ultraviolet and Optical Telescope (UVOT), to investigate the temporal and spectral characteristics of this enigmatic blazar. Particularly, we emphasized on exploring the variations in the broadband SED behavior as a function of the source activity with the primary motivation to understand the VHE emission and associated radiative mechanisms. Details of observations using the Fermi and Swift telescopes for the period and the data analysis are presented in Section 2. Section 3 presents the findings and Section 4 provides discussion on SED modeling. Section 5 discusses the findings and finally in section 6 we summarizes the study. We have adopted a flat cosmology with $\Omega_m = 0.3$, and $H_0 = 70$ km s$^{-1}$\,Mpc$^{-1}$.

\section{Multiwavelength Data Reduction}\label{sec:data}

\subsection{Fermi-Large Area Telescope}
 The Fermi-LAT, launched in 2008 June is a high-energy gamma ray space telescope. It images energies ranging from 20 MeV to energy greater than 300 GeV and covers 20 percent of the sky at any time \citep[][]{2009ApJ...697.1071A}. In this work we analyzed the data from 2023 November 12 to 2024 March 31  (MJD 60260-60400). The analysis was performed using \emph{fermipy}(Wood et.al.,2021) and the standard Fermi tools software. A circular region with $12^\circ$ radius around the source was chosen to extract the events with  $\tt{evclass=128}$ and $\tt{evtype=3}$ and the filters $\tt{DATA\_QUAL>0}$ and $\tt{LAT\_CONFIG==1}$ were applied as recommended by \emph{Fermi}-LAT documentation. 
The {${\tt{gll\_iem\_V07}}$} file along with the ${\tt{iso\_P8R3\_SOURCE\_V2\_v1}}$ files were used to model the diffuse background gamma-ray emissions. To reduce the contamination from the Earth limb gamma-ray, a zenith angle cut of $z_{\rm max}<90^{\circ}$ was chosen. The source model file was generated using the Fermi 4FGL-DR4 catalog (Abdollahi et al. 2020), including the sources within a circular region of interest (ROI)(12+10) degree. A maximum likelihood (ML) test statistics \mbox{TS = 2$\Delta$log(L)}, where L is the ratio of the likelihood values for models with and without a $\gamma$-ray point object, was used to determine the signifcance of $\gamma$-ray signal. Initially, the likelyhood analysis for the entire period was carried out keeping the spectral parameters of the sources within the 10 degree ROI free, while those for the sources outside the ROI were frozen to the catalog values. In the final model file, we fixed the spectral parameters for the background sources with TS$<$25, and then used it for generating the light curve. For the light curve analysis, the detection of the source is considered only if TS$>$9, i.e.,confidence level greater than $3\sigma$.
 
\subsection{Swift Observations}
During the period covering the VHE detection and the associated $\gamma$--ray flaring events, OP 313 was observed for 35 times by the Swift satellite. To carry out the X-ray spectral analysis, we analyzed the Swift-XRT data using the automated online tool $\tt{\emph{Swift}-XRT \ data\ product's\ generator}$ \citep{2009MNRAS.397.1177E}. This software automatically chooses the source and background regions depending on the source count rate and also consider the pile-up effects, if any. We generated the observation-binned, i.e., one flux point per observation, light curve using this tool. The X-ray spectra of the time periods selected for the SED modeling (details below) were also generated using the online tool. The X-ray spectra were then loaded in XSPEC \citep{1996ASPC..101...17A} and fitted with either a power-law or log-parabola model (depending on the fit quality) including the Galactic absorption. The absorption-corrected X-ray fluxes were estimated after fixing the neutral hydrogen column density to \mbox{N$_{\rm H}=1.24\times10^{20}$\ cm$^{-2}$} \citep{Kalberla_2005}. 

The UVOT observations were downloaded from the High Energy Astrophysics Science Archive Research Center and analyzed using standard procedures\footnote{\url{https://swift.gsfc.nasa.gov/analysis/threads}}. The tool \mbox{\emph{uvotimsum}} was used to sum the images in all the available filters. A circular region of radius 5 arcsec around the source was identified for extracting the counts, while another circle with 20 arcsec radius in a nearby source-free region was used for background estimation. Aperture photometry was performed using the tool \emph{uvotsource} for measuring fluxes in each filter. The fluxes were corrected for Galactic extinction by fixing \mbox{E(B-V)=\,0.0125 for R$_V$=A$_V$/E(B-V)=3.1} \citep{Schlafly_2011}. 
\section{Results}\label{sec:res}

\begin{figure*}
    \centering
    \includegraphics[scale=0.7]{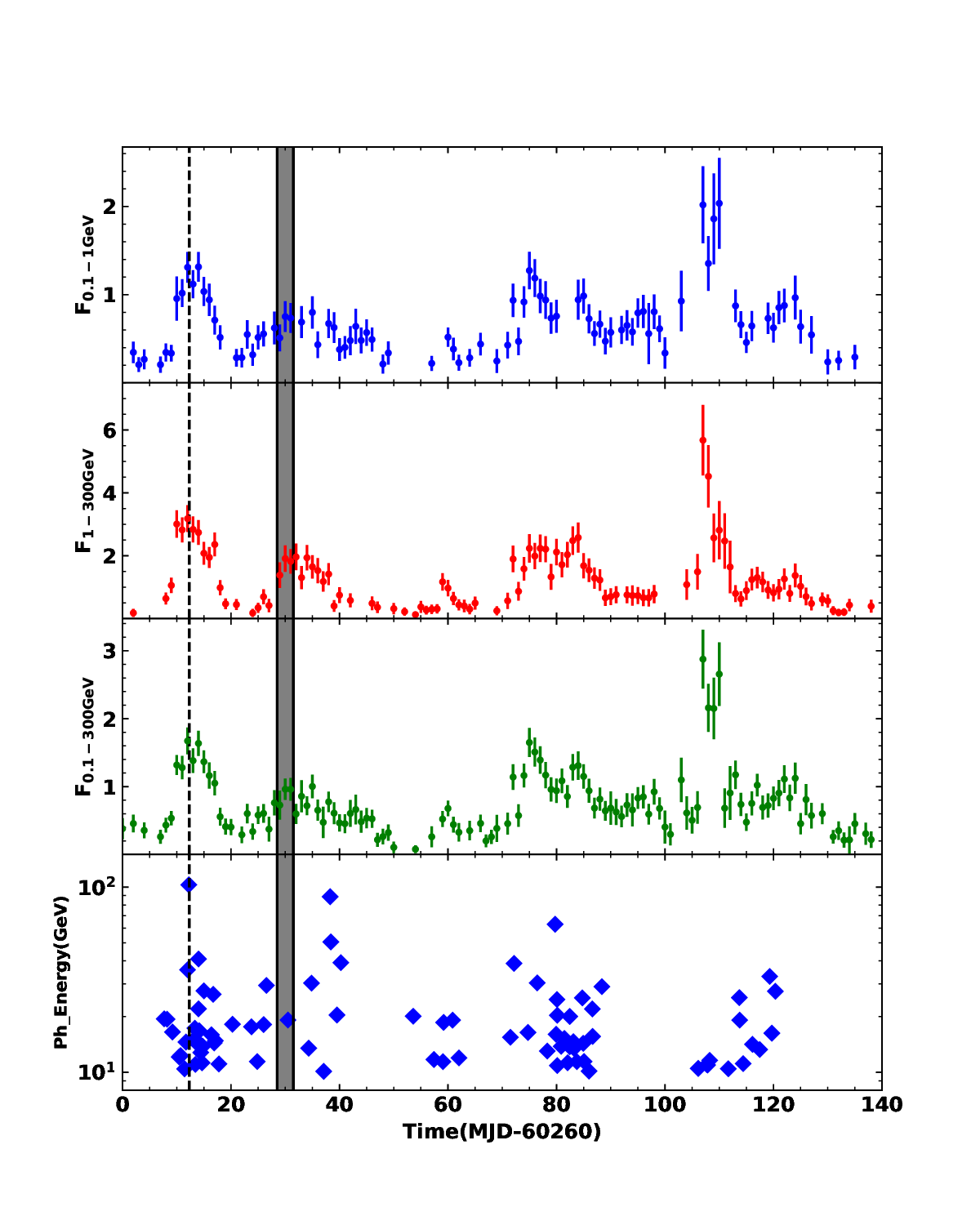}
    \caption{One day binned \gm-ray lightcurve of OP 313 covering the time period MJD 60260-60400. The data points in the \mbox{0.1$-$1 GeV} and 0.1$-$300 GeV energy band panels are in units of $10^{-6}$ ph\ cm$^{-2}$\ s$^{-1}$. The 1$-$300 GeV band light curve is in units of \mbox{$10^{-7}$ ph\ cm$^{-2}$\ s$^{-1}$.} The bottom panel shows the energy of the highest energy photons detected during the covered time period. Dotted vertical line shows the time when Fermi detected VHE photons. On the other hand, the shaded region represents the epoch of the VHE detection by LST-1.}
    \label{gamma_lightcurve}
\end{figure*}


\subsection{Gamma-ray Variability}
We generated the one-day binned \gm-ray light curves of OP 313 in three energy bands: 0.1$-$300 GeV, 0.1$-$1 GeV, and 1$-$300 GeV, as shown in Figure \ref{gamma_lightcurve}. Furthermore, the energy of the highest energy photons in each time bin was estimated using the tool {\tt gtsrcprob}. In Figure~\ref{gamma_lightcurve}, the bottom panel shows the epochs were photons with energy greater than 10 GeV are detected with source probability greater than 95 percent. Several episodes of flaring activity can be seen, and we highlight the epoch of VHE detection with LST-1 and also the time when Fermi-LAT detected photon with energy grater than 100 GeV photon from OP 313. The Fermi-LAT detected a photon of energy 102.8 GeV on MJD 60273 with the source probability of 99\%. Interestingly, the VHE photon was detected with the Fermi-LAT at the peak of a \gm-ray flare, whereas, the VHE detection with LST-1 was made when OP 313 was in a relatively lower \gm-ray activity state. A \gm-ray outburst of high amplitude was observed around MJD 60370.

\begin{figure}
    \centering
    \includegraphics[width=1\linewidth]{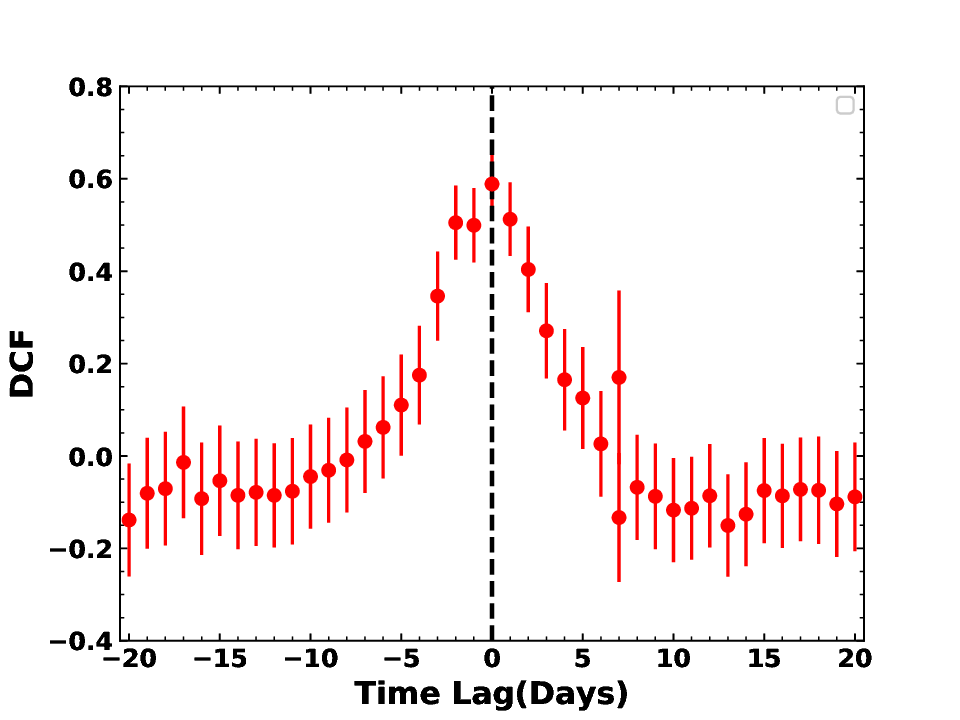}
    \caption{Discrete Correlation Function (DCF) calculated for the 0.1\,\text{--}\,1\ \text{GeV} and 1\,\text{--}\,300\ \text{GeV} light curves. The time ordering is \mbox{T(1\,\text{--}\,300\ \text{GeV})\text{--}T(0.1\,\text{--}\,1\ \text{GeV}).}}
    \label{DCF}
\end{figure}

The presence/absence of a time lag between low- and high-energy \gm-ray emission could provide clues about the location of \gm-ray emitting region, i.e., inside or outside the BLR \citep[][]{2012ApJ...758L..15D}. Therefore, we investigated the presence/absence of any significant time lead/lag in the 0.1$-$1 GeV and 1$-$300 GeV light curves by utilizing the \textit{z-transformed discrete correlation function} \citep[ZDCF method;][]{2013arXiv1302.1508A}. The outcome of this exercise is displayed in Figure~\ref{DCF}. The uncertainties were computed using a Monte Carlo simulation. 
We did not find any significant lead/lag between the two energy bands which could be intrinsic to the source. Alternatively, it could also be due to the limited time resolution of the light curves and the proximity of the LAT energy bands.

\begin{figure*}
\centering
    \includegraphics[width=\linewidth]{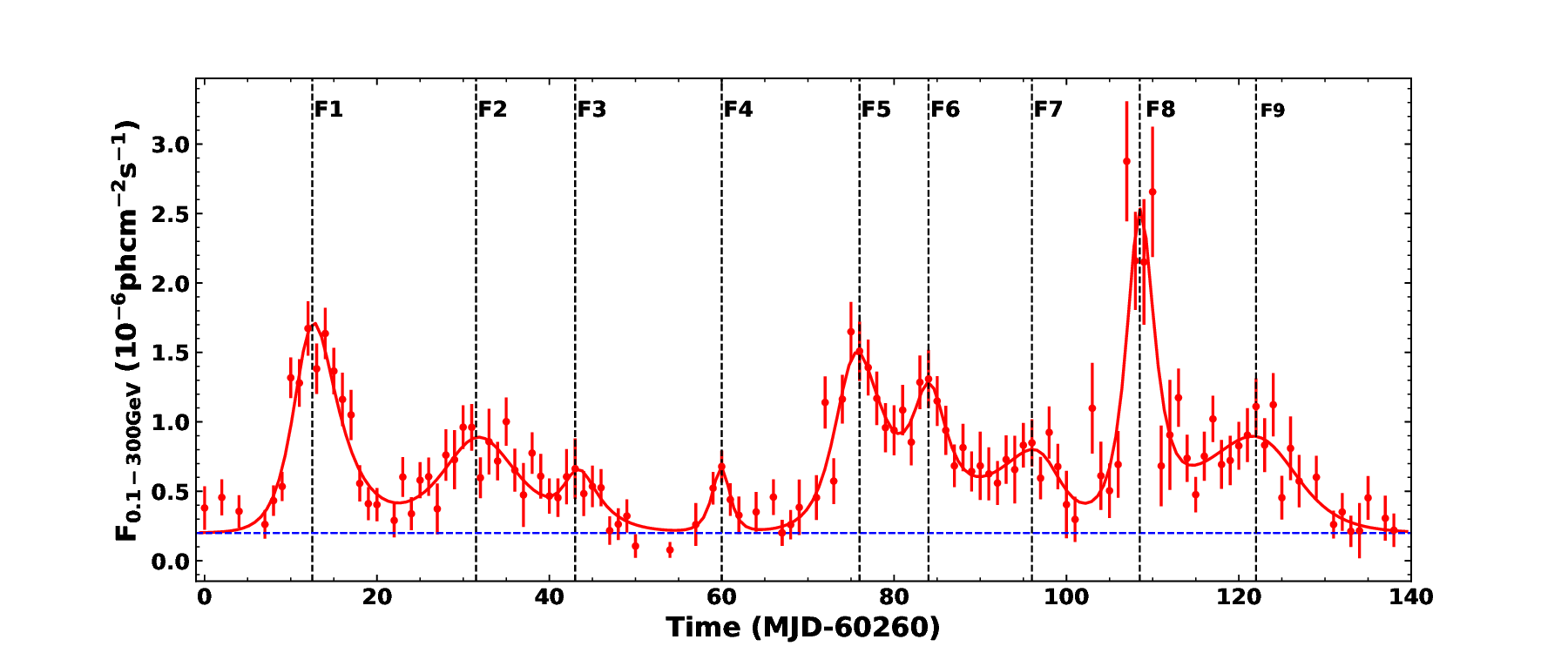}

\caption{One day binned $\gamma$-ray light curve of OP 313 fitted with exponential function defined in equation~\ref{exponentialfit}.}
 \label{fitted_gammalightcurve}
\end{figure*}

The visual inspection of  the 0.1$-$300 GeV light curves led to the identification of nine flares, denoted as F1-F9 and are shown in Figure \ref{fitted_gammalightcurve}. The profiles of these flares, e.g., flare rise and decay time, were analyzed by fitting an exponential function \citep[][]{2010ApJ...722..520A} represented by the following equation 

\begin{equation}
\label{exponentialfit}
    F(t)=F_{b}+2F_{0}\left[\textrm{exp}\left(\frac{T_{0}-t}{T_{r}}\right)+\textrm{exp}\left(\frac{t-T_{0}}{T_{d}}\right)\right]
 \end{equation}
where, $F_{b}$ is the baseline flux, $T_{r}$ and $T_{d}$ are the flare rise and decay time, respectively. The parameter $F_{0}$ is the flux at time $T_{0}$ representing approximately the flare amplitude. The asymmetry parameter $\zeta$ was calculated using the following equation
\begin{equation}
    \label{asymetry}
    \zeta=\frac{T_{d}-T_{r}}{T_{d}+T_{r}}.
\end{equation}
The fitted function along with one day binned light curve is shown in Figure \ref{fitted_gammalightcurve}. The obtained parameters are tabulated in Table \ref{rise_falling time}. For the third flare we could not fit the equation~\ref{exponentialfit}. Hence, for this flare the best fit parameters are quoted without errors in Table \ref{rise_falling time} with suffix *. 
%
We also scanned the light curve data to determine the shortest timescale of flux variability using the following equation \citep[e.g.,][]{2011A&A...530A..77F}
\begin{equation}
     F(t) = F(t_0)\,2^{{-(t-t_0)}/t_{\rm var}}
     \label{eqn:flux doubling}
\end{equation}
where $F(t)$ and $F(t_0)$ are the fluxes at two consecutive times $t$ and $t_0$. We found three time periods in which the estimated flux doubling/halving time was at least 3$\sigma$ significant (Table~\ref{tab:halving doubling time}). The fastest flux variation were observed in the time bin MJD 60366-60367 with the flux doubling time of  \mbox{11.7\ $\pm$\ 4.4} hours.

\begin{table*}
\caption{Flare characteristics obtained from the temporal profile fitting of one day binned $\gamma$-ray light curve. The flare amplitude $F_{0}$ is measured in units of $10^{-6}$ ph\ cm$^{-2}$\ s$^{-1}$.}
\setlength{\tabcolsep}{20pt}
\renewcommand{\arraystretch}{1.2}
\begin{tabular}{cllllr}
\hline
\textbf{Flare Name}&\textbf{$T_{0}$(MJD)} & \textbf{\ \ \ \ $F_{0}$} & \textbf{$T_{r}$(day)} & \textbf{$T_{d}$(day)} &\textbf{$\zeta$\ \ \ }\\
\hline
F1 & 60272.5     &1.69$\pm$0.16        &3.09$\pm$0.32        &1.44$\pm$0.10      &0.29    \\
F2 & 60292.5     & 0.68$\pm$0.07       & 3.71$\pm$0.68       & 4.46$\pm$0.68     &0.09\\
F3 & 60303.5     & 0.35*               & 1.54*               & 1.72*             &0.05  \\
F4 & 60320.5     & 0.47$\pm$0.11       & 0.97$\pm$0.45       & 0.89$\pm$0.36     &-0.04 \\
F5 & 60336.7     & 1.20$\pm$0.13       & 1.78$\pm$0.32       & 3.22$\pm$0.81     &0.29  \\
F6 & 60344.7     & 0.79$\pm$0.16       & 1.62$\pm$0.70       & 1.77$\pm$0.62     &0.04  \\
F7 & 60356.7     & 0.43$\pm$0.09       & 7.85$\pm$3.56       & 1.69$\pm$0.89     &-0.64\\
F8 & 60369.7     & 2.11$\pm$0.30       & 1.34$\pm$0.25       & 1.45$\pm$0.32     &0.04\\
F9 & 60383.0     & 0.62$\pm$0.08       & 8.90$\pm$2.54       & 3.25$\pm$0.67     &-0.46\\
\hline
&$\chi_{red}^2$=1.26  &    dof=98 \\
\hline
\end{tabular}
\label{rise_falling time}
\end{table*}

\begin{table*}
	\centering
	\setlength{\tabcolsep}{12pt}
	\setlength\extrarowheight{4pt}
        \caption{Details of the fast flux doubling/halving times estimated from the analysis. The parameters t$_0$ and t$_1$ are the times in MJD; F(t$_0$) and F(t$_1$) are the fluxes in units of 10$^{-6}$ ph cm$^{-2}$ s$^{-1}$; significance of difference in flux in $\sigma$; variability time scale ($t_{\rm var}$) in hours.}
	\label{tab:tvar}
	\begin{tabular}{llllccc}
		\hline 
		t$_0$ & t$_1$ & F(t$_0$) & F(t$_1$) & Significance & $t_{\rm var}$ (hour) \\
		\hline\hline
60269 	 & 60270 	 & $ 0.54 \pm 0.11 $ 	 & $ 1.32 \pm 0.14 $ 	 & 4.4 	 & $18.42\pm3.4$ \\
60366 	 & 60367 	 & $ 0.70 \pm 0.24 $ 	 & $ 2.90 \pm 0.40 $ 	 & 4.4 	 & $11.68\pm4.4$ \\
60370 	 & 60371	 & $ 2.66 \pm 0.50 $ 	 & $ 0.69\pm 0.30 $ 	 & 3.6 	 & $12.24\pm4.9$ \\
		\hline
	\end{tabular}
 \label{tab:halving doubling time}
\end{table*}


\subsection{X-ray spectral Variability}
\begin{figure*}
    \centering
    \includegraphics[width=1\textwidth]{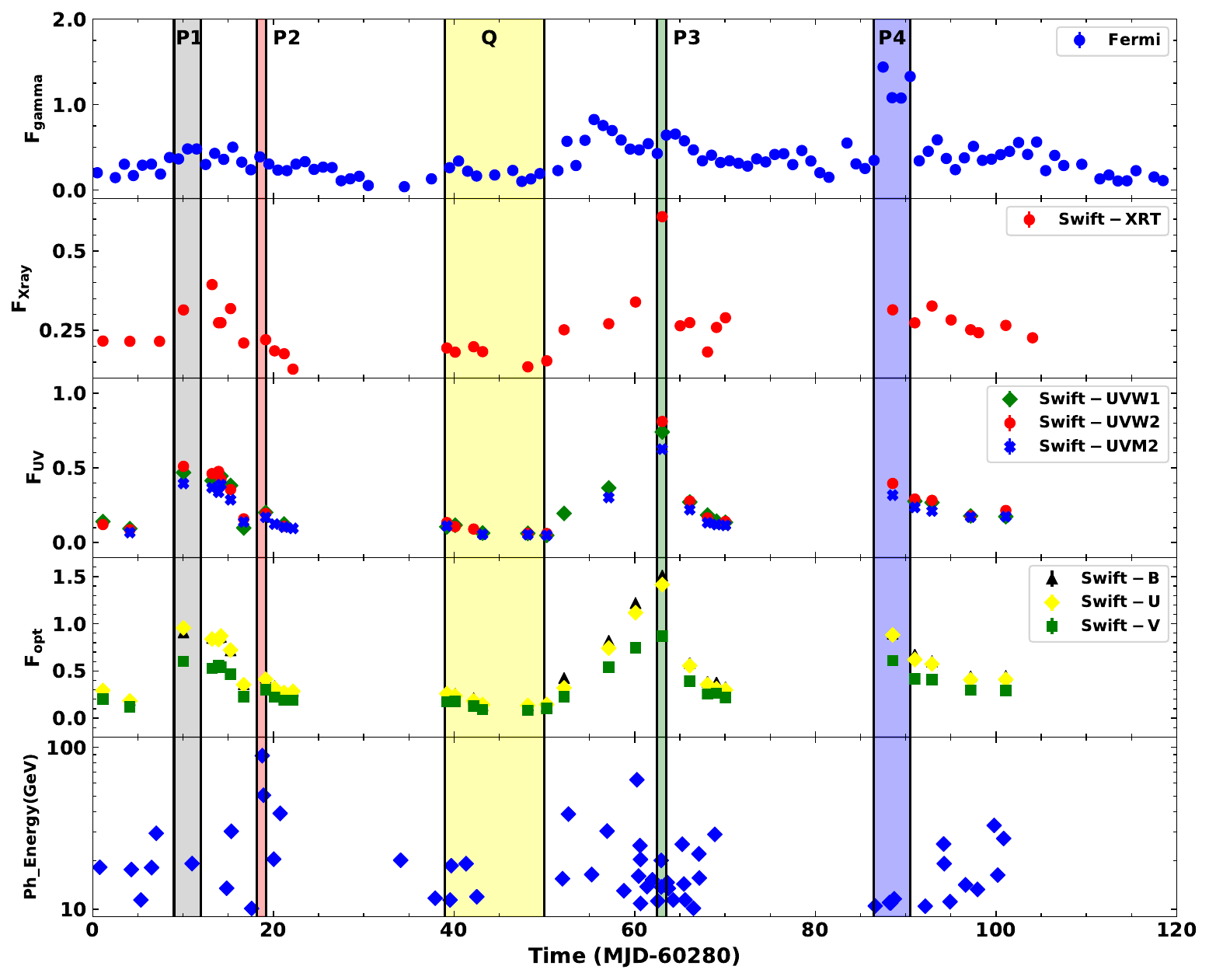}
    \caption{Multi-wavelength light curves of OP 313. One day binned Fermi\text{--}Lat data points are in units of $10^{-6}$ ph\ cm$^{-2}$\ s$^{-1}$, Swift\text{--}XRT, Swift\text{--}UV and Swift\text{--}Optical data points are in units of $10^{-11}$ erg\ cm$^{-2}$\ s$^{-1}$. The colored shaded regions highlight the epochs selected for broadband SED generation and modeling.}
    \label{multi-wavelength lightcurve}
\end{figure*}

We also looked for the activity in lower energy bands corresponding to the flaring episodes in \gm-ray. Figure \ref{multi-wavelength lightcurve} shows the multi-wavelength light curve for the period MJD 60280-60390 using the data collected from the Fermi-LAT and Swift satellites. From the multi-wavelength light curve, it is evident that the source showed flux variations across the optical-UV, X-ray, and \gm-ray bands. Interestingly, a comparison of the X-ray and \gm-ray light curves revealed an intriguing variability behavior. The brightest X-ray flare was observed around MJD 60345 and the source was in a moderately bright \gm-ray flux state. On the other hand, a brighter \gm-ray flaring episode was observed around MJD 60370; however, the X-ray flux activity was relatively moderate. To understand the underlying radiative processes responsible for the observed patterns, we generated the broadband SED for several epochs as described below.


We chose five different epochs from multi-wavelength light curve for the SED generation and modeling. This included the epoch of VHE detection with LST-1. There were no near-simultaneous multi-frequency observations taken at the time of the detection of the 102 GeV, i.e., VHE photon with the Fermi-LAT. Therefore, we considered another epoch when the second highest energy photon of 88 GeV was detected. For the simplicity, we call this period as the Fermi-LAT VHE detection period. We also considered the brightest X- and \gm-ray flaring episodes, and a quiescent activity period. The details of epochs with covered time periods and acronym are shown in Table \ref{epochs} and are highlighted in Figure~\ref{multi-wavelength lightcurve}. To minimize the averaging of the Fermi-LAT data that can washout interesting spectral features, we selected only one day binned LAT data centered at the time of Swift pointing for the X-ray flare and Fermi-LAT VHE detection epochs. On the other hand, LST-1 VHE detection epoch was chosen to be three day long to remain consistent with the LST-1 observation. We selected four day binned LAT data to fully cover the brightest \gm-ray flaring episode. On the other hand, the low-activity state was chosen to be long enough to have meaningful \gm-ray photon statistics. The choice of selection of epochs also depended on the availability of observations in all energy bands. 

\begin{figure}
    \includegraphics[width=1\linewidth]{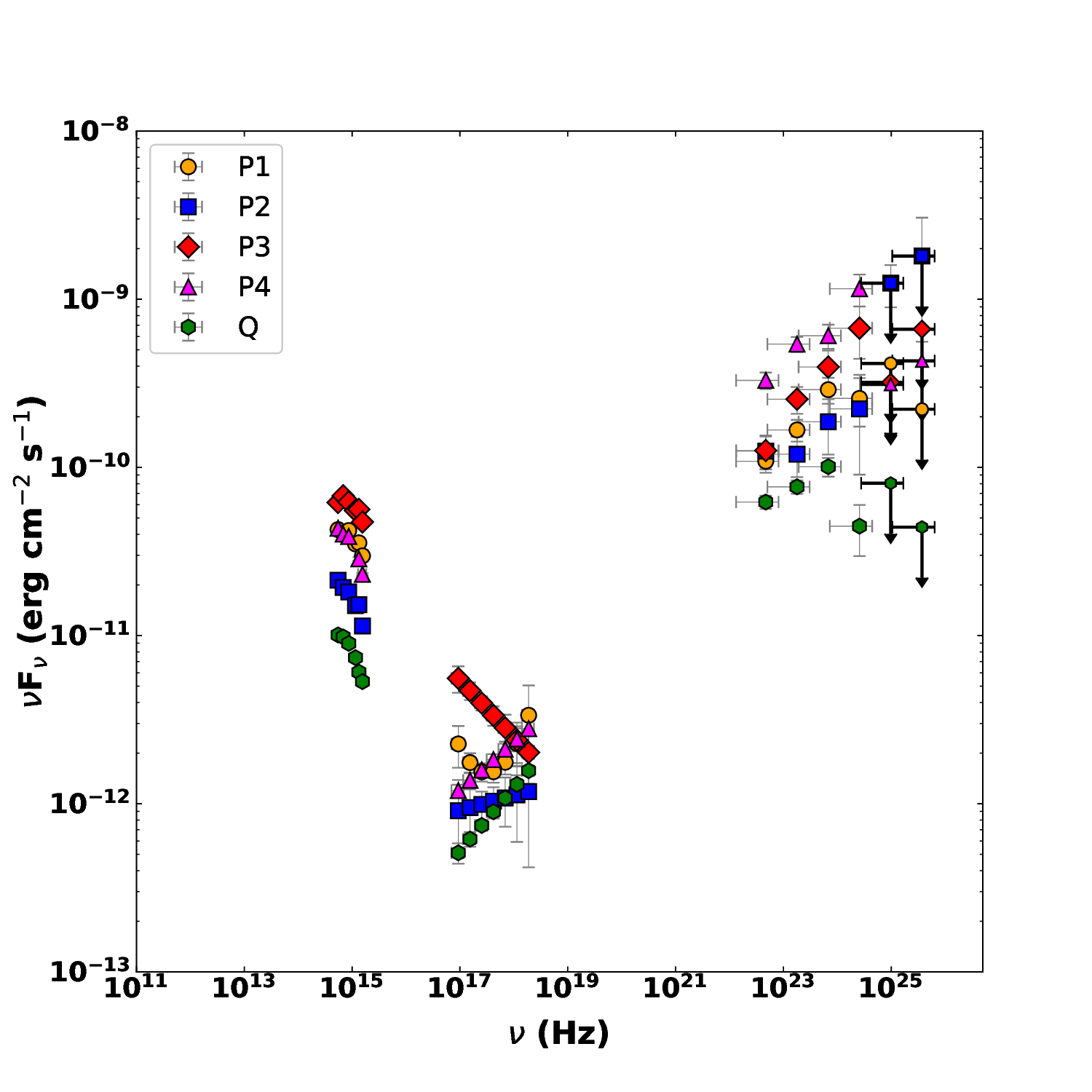}
    \caption{Combined plot of all the broadband SEDs.}
    \label{combined_SED}
\end{figure}

\begin{table}
\caption{Different epochs selected for the broadband SED generation and modeling.}
\setlength{\tabcolsep}{5pt}
\renewcommand{\arraystretch}{1.2}
\begin{tabular}{llc}
    \hline
\textbf{Epoch}          & \textbf{Time (MJD)}& \textbf{Acronym} \\
    \hline
LST-1 Detection           & 60289\text{--}60292       & P1      \\
Fermi-LAT VHE detection & 60298.2\text{--} 60299.2 & P2      \\
Brightest X-ray flare   & 60342.5\text{--} 60343.5 & P3      \\
Brightest \gm-ray flare         & 60366.5\text{--}60370.5   & P4      \\
Quiescent state         & 60319\text{--}60330       & Q       \\
\hline
    \end{tabular}
    \label{epochs}
\end{table}
The generated SEDs corresponding to the selected epochs are shown in Figure \ref{combined_SED} and we provide the corresponding spectral parameters in Table~\ref{powerlaw spectral index} and \ref{tab:swift}. A close inspection of Figure~\ref{combined_SED} revealed that the optical-UV spectrum primarily exhibited flux variations while maintaining its spectral shape. Similar observations were made in the \gm-ray spectrum which was found to exhibit a flat rising shape. The most interesting spectral behavior was identified in the X-ray band where \mbox{OP 313} not only showed flux variations but also a considerable spectral evolution. For example, it had a hard X-ray spectrum during the low- and the brightest \gm-ray activity states (epoch Q and P4 in Figure~\ref{combined_SED}), similar to that typically observed from FSRQs \citep[e.g.,][]{2015ApJ...803...15P}. During the epoch of LST-1 VHE detection, the X-ray spectrum showed a concave upward feature indicative of the presence of a break likely due to falling synchrotron emission at soft X rays and rising inverse Compton radiation at higher energies. Similar X-ray spectral behavior were also noticed during flaring episodes of other FSRQs \citep[][]{2013MNRAS.432L..66G}. The concave upward spectral shape disappeared during the brightest X-ray flaring period where a steep falling X-ray spectrum was observed. Finally, OP 313 exhibited a soft X-ray spectrum during the Fermi-LAT VHE detection epoch. To our knowledge, this is probably the first time when such drastic spectral changes were seen in the X-ray spectrum of a FSRQ within a few months thereby indicating the dynamic nature of the source. 

\begin{table*}
    \centering
	\setlength{\tabcolsep}{10pt}
	\caption{Best fit power law spectral index to the Fermi-LAT observations. Integrated $\gamma$-ray flux(0.1-300 GeV) is in units of \mbox{$10^{-7}$ ph\ cm$^{-2}$\ s$^{-1}$.}}
    \begin{tabular}{lr>{\raggedleft\arraybackslash}ccc}
        \hline
        \multicolumn{4}{c}{\emph{Fermi}-LAT} \\
        \hline
        \text{Epoch} &\hspace{5pt} \text{Flux (0.1-300 GeV)} & \text{Photon Index}& \text{TS} \\
        \hline
        P1& 8.55$\pm$0.04 \ \ \ \          & 1.78$\pm$0.06           & 340.90      \\
        P2& 7.29$\pm$0.04  \ \ \ \         & 1.77$\pm$0.01           & 216.43      \\
        P3& 11.58$\pm$0.17  \ \ \ \        & 1.68$\pm$0.08           & 406.76      \\
        P4& 25.74$\pm$0.23  \ \ \ \        & 1.81$\pm$0.05           & 763.90      \\
        Q & 3.98$\pm$0.03   \ \ \ \        & 1.94$\pm$0.05            &  205.24      \\
\hline
    \end{tabular}
    \label{powerlaw spectral index}
\end{table*}
\begin{table*}
	\centering
	\setlength{\tabcolsep}{15pt}
	\caption{The summary of the data analysis performed for the Swift XRT and UVOT observations.In bottom panel U, B, V, UVW1, UVM2, UVW2 represents fluxes in Visual, Blue, UltraViolet, Ultraviolet Wide 1, Ultraviolet Medium 2,Ultraviolet Wider 2  filters respectively. Fluxes are in units of $10^{-12}$ ergs cm$^{-2}$ s$^{-1}$}
	\label{tab:swift}
	
	\begin{tabular}{lcccc}
         
		\hline
		\hline
		\multicolumn{5}{c}{\emph{Swift}-XRT} \\
		\hline
		Epoch & Flux & Photon Index ($\Gamma$/$\alpha$) & Curvature ($\beta$) & $\chi^{2}$/Dof \\
		\hline
		P1 & $(7.3 \pm 1.4) $ & $2.74 \pm 0.50$ & $-0.60\pm 0.30$ & 47.77/48 \\
		P2 & $(3.6\pm 0.9)$ & $1.94 \pm 0.40$ & - & 19.38/24 \\
		P3 & $(12.1\pm3.3)$ & $2.40 \pm 0.20$ & - & 48.19/47 \\
        P4 & $(5.1\pm0.5)$ & $1.80 \pm 0.10$ & - & 99.24/124 \\ 
        Q  & $(3.0\pm0.3)$ & $1.63 \pm 0.10$ & - & 80.72/107 \\
		\hline
	\end{tabular}
    \setlength{\tabcolsep}{6.5pt}
	\vspace{3pt} 
	\begin{tabular}{lcccccc}
		\multicolumn{7}{c}{\emph{Swift}-UVOT}\\
		\hline
		&V &B& U& UVW1& UVM2& UVW2\\
		\hline
		P1 & 1.65 $\pm$ 0.04 & 2.00 $\pm$ 0.03 & 1.67 $\pm$ 0.03 & 1.23 $\pm$ 0.02 & 0.89 $\pm$ 0.02 & 0.99 $\pm$ 0.02\\
		P2 & 0.83 $\pm$ 0.03 & 0.95 $\pm$ 0.02 & 0.72 $\pm$ 0.02 & 0.53 $\pm$ 0.01 & 0.38 $\pm$ 0.01 & 0.38 $\pm$ 0.01 \\ 
		P3 & 2.40 $\pm$ 0.05 & 3.34 $\pm$ 0.06 & 2.47 $\pm$ 0.05 & 1.97 $\pm$ 0.04 & 1.41 $\pm$ 0.03 & 1.57 $\pm$ 0.02 \\
		P4 & 1.68 $\pm$ 0.04 & 1.97 $\pm$ 0.03 & 1.54 $\pm$ 0.03 &  & 0.72 $\pm$ 0.02 & 0.77 $\pm$ 0.01 \\
		Q  & 0.39 $\pm$ 0.02 & 0.48 $\pm$ 0.02 & 0.36 $\pm$ 0.01 & 0.26 $\pm$ 0.01 & 0.15 $\pm$ 0.01 & 0.18 $\pm$ 0.01 \\ 
		\hline
        \hline
	\end{tabular}

\end{table*}


\section{SED Modeling}\label{sec:SED}

To further explore the multi-wavelength spectral variability, we performed a systematic broad band SED modeling under leptonic emission scenario. We considered a homogeneous spherical blob with a radius $R$ embedded with a tangled magnetic field $B$ as the emission region. This blob moves down the blazar jet at a relativistic speed with a bulk Lorentz factor $\Gamma$ at an angle $\theta$ with respect to the observer's line of sight. The emission region is assumed to be populated by an electron distribution following a broken power-law:

\begin{align} \label{eq:broken}
	N(\gamma)\,d\gamma = \left\{
	\begin{array}{ll}
		K\,\gamma^{-p}\,d\gamma&\textrm{for}\quad \mbox {~$\gamma_{\rm min}<\gamma<\gamma_b$~} \\
		K\,\gamma_b^{q-p}\gamma^{-q}\,d\gamma&\textrm{for}\quad \mbox {~$\gamma_b<\gamma<\gamma_{\rm max}$~}
	\end{array}
	\right.
\end{align}
Here, $K$ is the normalization of the particle number density, $\gamma_{\rm min}$ and $\gamma_{\rm max}$ are the minimum and maximum energy of the electrons,respectively and $p$ and $q$ are the indices of the broken-power law particle distribution with $\gamma_b$ as the break energy. The electron distribution undergoes energy losses 
through synchrotron and inverse Compton processes. The source of seed photons for the IC process can be either synchrotron photons (SSC) or 
photons external to the jet (EC). The dominant sources of external photons are thermal IR photons from the dusty torus (EC-torus) and Lyman-alpha line 
emission from the broad line region (EC-BLR). The thermal emission from torus is assumed as black body peaking at 1000K. For numerical simplicity we assumed BLR emission as again black body with peak frequency equal to the Lyman-alpha emission line.

The numerical code capable of computing the emissivities resulting from synchrotron, SSC, and EC processes was added as a local model in XSPEC and used to fit the broadband SED of the source for the selected epochs. The main set of parameters that govern the broadband SED are $K$, $\gamma_{\rm min}$, $\gamma_{\rm max}$, $\gamma_{\rm b}$, $p$, $q$, $B$, $R$, $\Gamma$, $\theta$ 
and density of the external target photon field ($U_{ph}$).  We also introduced an additional equipartition parameter $\eta$ which is the ratio between the particle
energy density to the magnetic field energy density.  Significant deviation of $\eta$ from unity indicates the emission region is not in equilibrium and can be 
unstable \citep[][]{1970ranp.book.....P}.

\begin{figure}
\centering
\includegraphics[width=0.9\columnwidth]{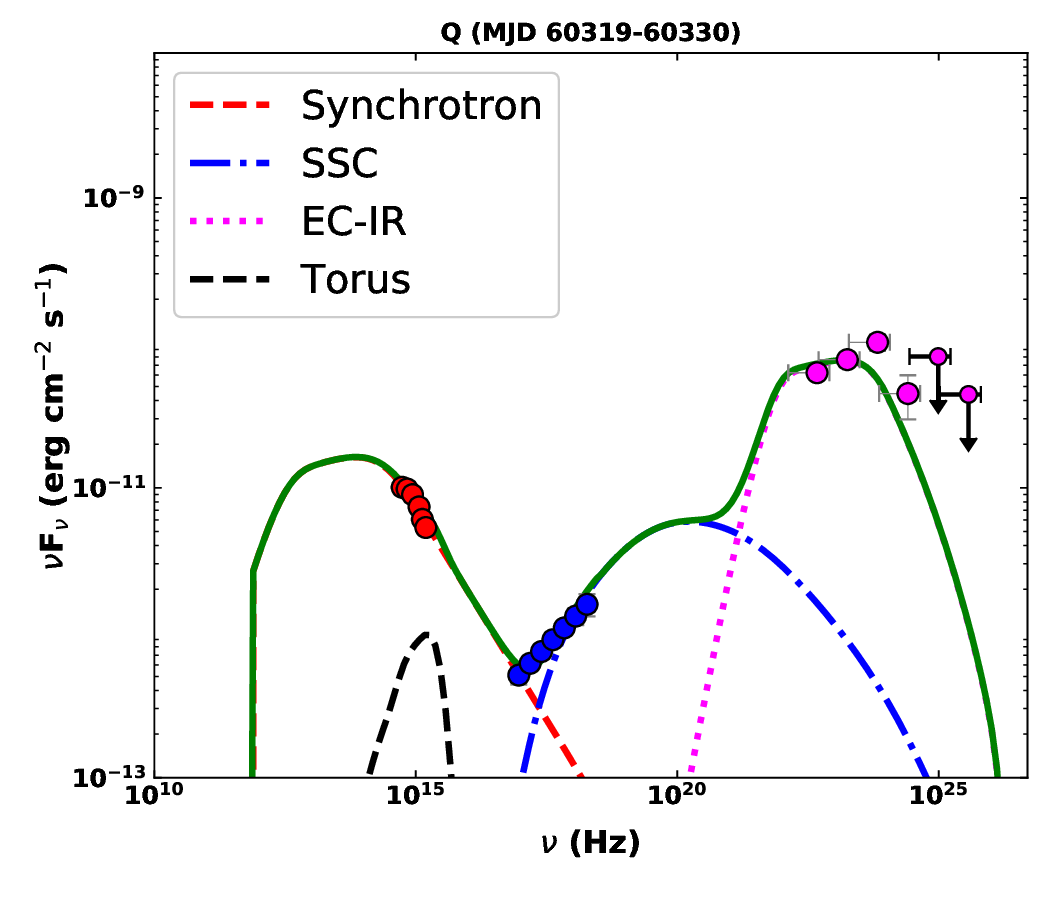}
		
\caption{Plot showing result of modeling of the multi-wavelength SED for quiescent state. The red, blue, and magenta data points correspond to the Swift-UVOT, XRT and Fermi-LAT observations. The black dashed line refers to the blackbody emission from the dusty torus. On the other hand, red dashed, blue dash-dot, and magenta dotted lines show the synchrotron, SSC, and EC-torus models respectively. The sum of all the radiative components is shown with the green solid line.}
	\label{sed_low}	
\end{figure}

\begin{figure*}
	\begin{subfigure}{0.5\textwidth}
		\includegraphics[width=0.85\textwidth]{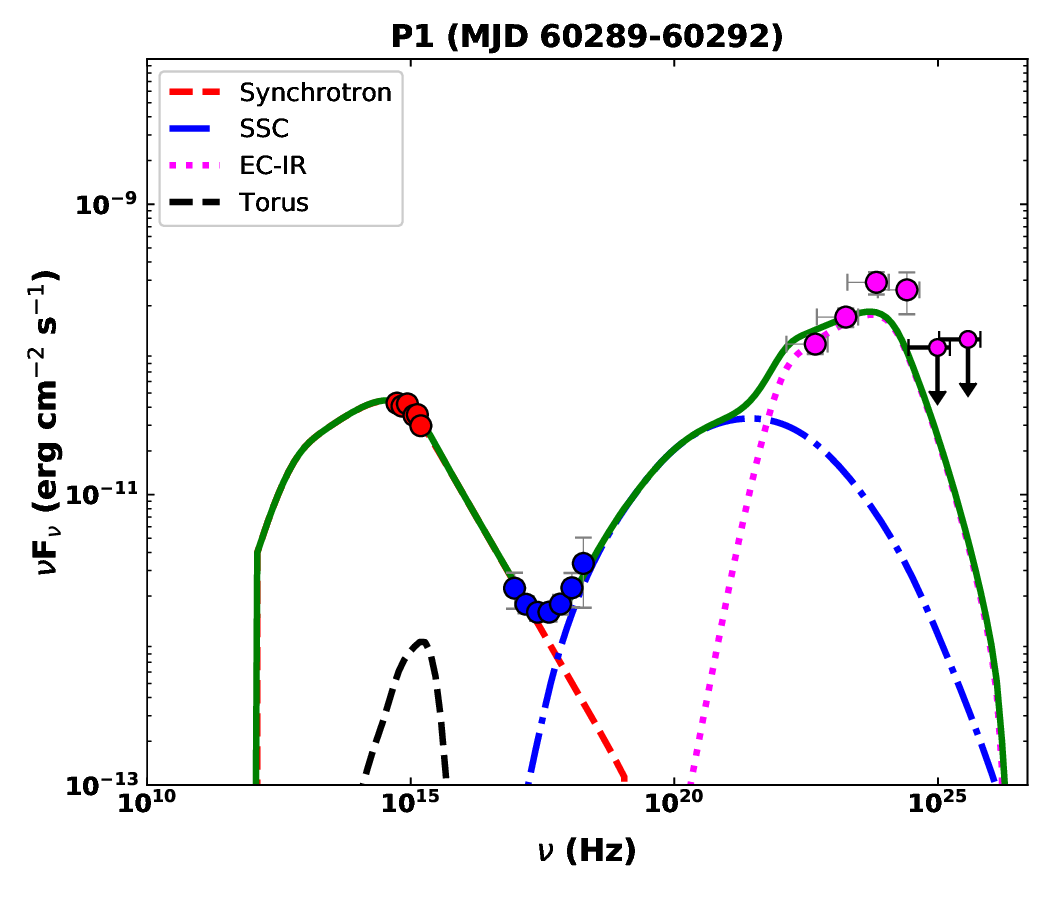}
	\end{subfigure}
	\hfill
	\hspace{-2cm}
	\begin{subfigure}{0.5\textwidth}
		\includegraphics[width=0.85\textwidth]{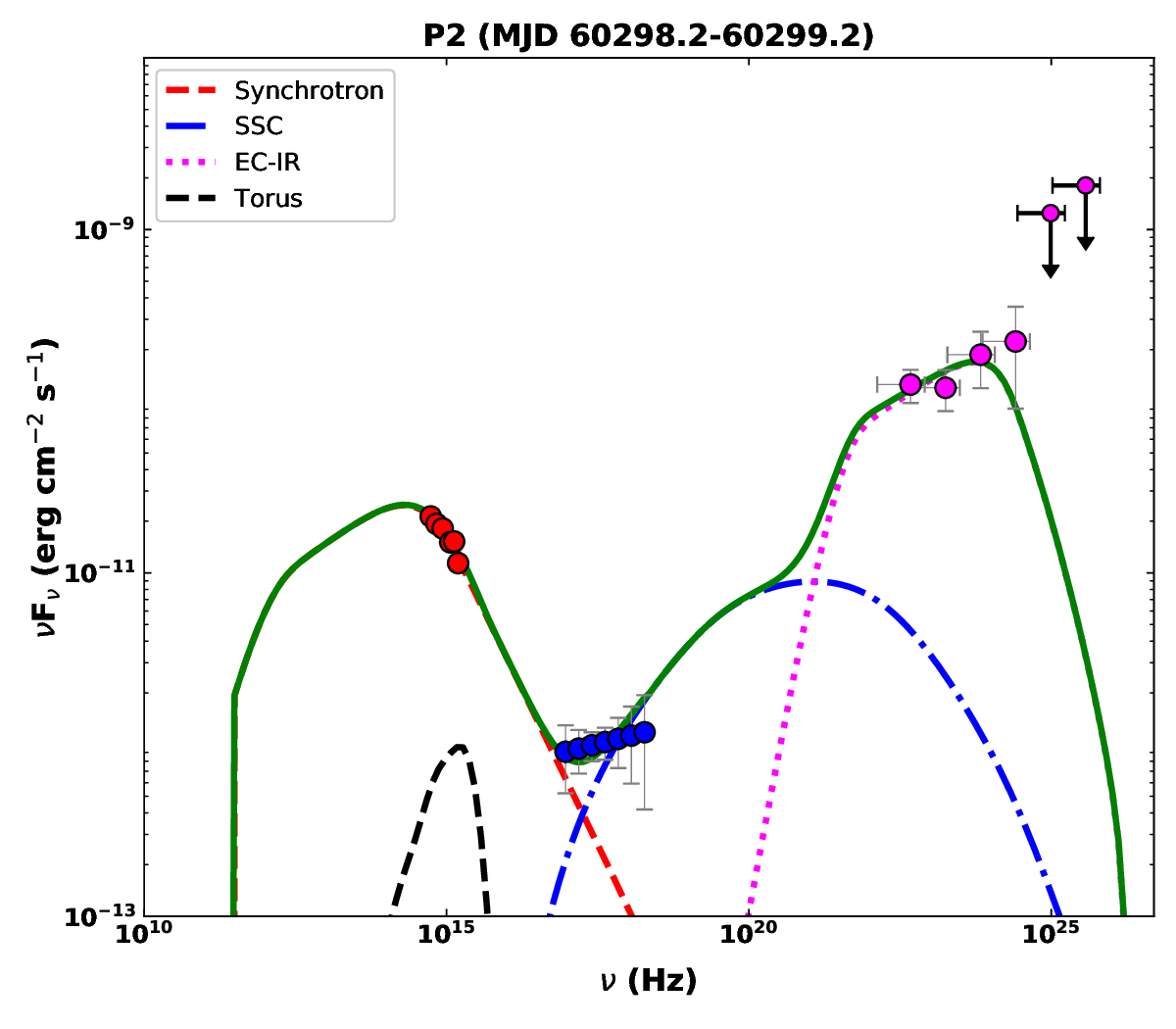}
	\end{subfigure}
    \begin{subfigure}{0.5\textwidth}
		\includegraphics[width=0.85\textwidth]{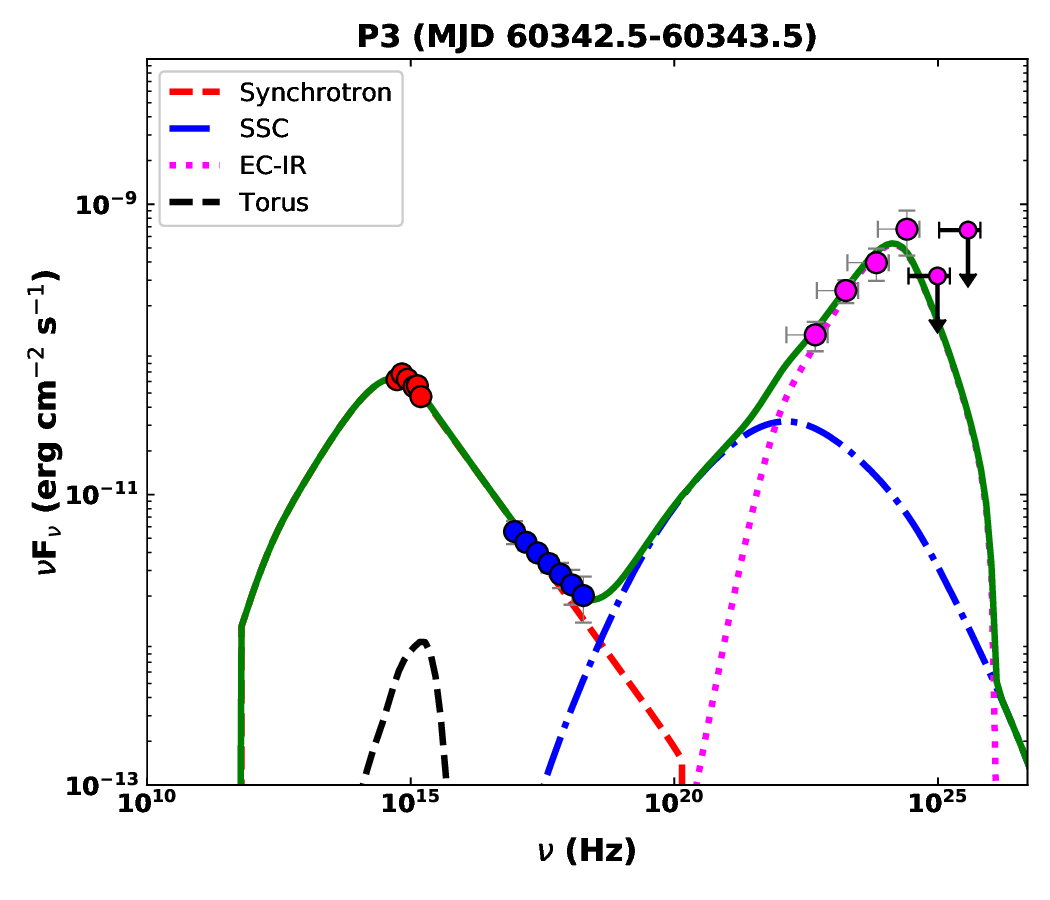}
	\end{subfigure}
	\hfill
	\hspace{-2cm}
            \begin{subfigure}{0.5\textwidth}
		\includegraphics[width=0.85\textwidth]{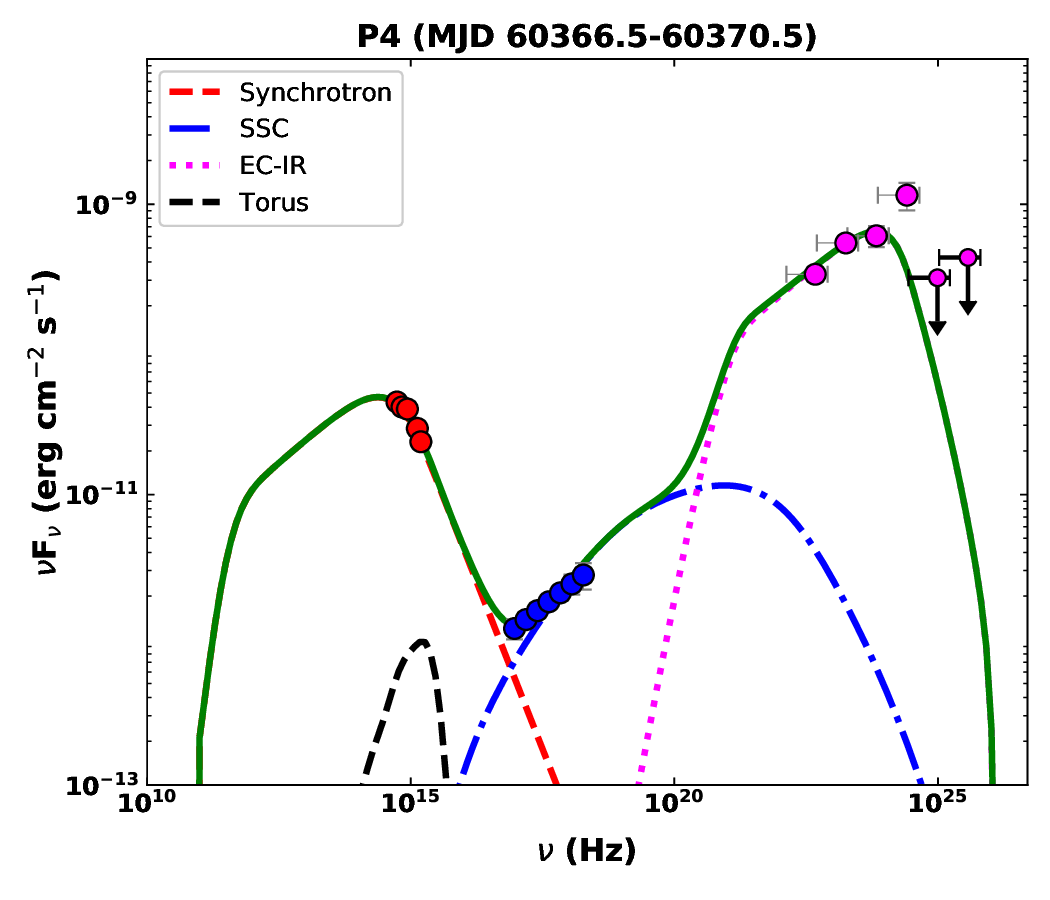}
	\end{subfigure}
 \caption{Same as Figure~\ref{sed_low} but for SEDs covering different activity states.}
 \label{sed_4 epochs}
 \end{figure*}

The initial fit was performed for the low flux state (Figure~\ref{sed_low}) with most of the parameters set free to vary. However, due to the
limited information available from the optical/UV, X-ray and $\gamma$ -ray bands, all parameters cannot be constrained and confidence
intervals are obtained only for $p$, $q$, $B$, $\Gamma$, $\gamma_b$ and $U_{ph}$. The rest of the parameters are fixed to their typical/best-fit values, and the fitting was performed. With the parameter set obtained for the low-flux state as an initial guess, the SEDs of the other flux states were performed. The SEDs of the best-fit model along with the observed data are shown in Figures~\ref{sed_low} and \ref{sed_4 epochs}. 
The model parameters corresponding to this fit are tabulated in Table \ref{tab:sed}. We found that the model with synchrotron, SSC, and EC-torus emission mechanisms reasonably explains the observed fluxes. It can be noted that the unusual spectral variability at the X-ray band is associated with the relative dominance of the particular emission component at this energy.


    \label{combined modelled sed}

\begin{table*}
\centering
\setlength\tabcolsep{10pt}
\caption{Best fit values of the model parameters from broadband SED fitting.}
 	\label{tab:sed}
 	\setlength{\tabcolsep}{15pt}
 	\begin{tabular}{ccccccC}
 		\hline
 		\hline
Parameters &Q	&	P1	&	P2	&	P3	& P4\\
\hline
\hline 
\\
 $p$	&	2.75$^{+0.16}_{-0.12}$		&	2.47$^{+0.14}_{-0.23}$	&	2.48$^{+0.18}_{-0.27}$	&	1.71$^{+0.12}_{-0.50}$ &  2.36$^{+0.05}_{-0.1}$	\\
 \\
$q$	&	4.17$^{+0.08}_{-0.06}$		&	4.26$^{+0.08}_{-0.07}$	&	4.44$^{+0.59}_{-0.27}$	&	4.01$^{+0.04}_{-0.03}$	& 4.84$^{+0.29}_{-0.15}$	\\
\\
 $\gamma_b$	&	2362$^{+578}_{-289}$		&	3923$^{+285}_{-247}$	&	4177$^{+982}_{-869}$	&	3537$^{+192}_{-314}$ &	3077$^{+198}_{-249}$	\\
 \\
$B$ 	&	0.44$^{+0.01}_{-0.01}$		&	0.51$^{+0.03}_{-0.04}$	&	0.28$^{+0.02}_{-0.01}$	&	0.34$^{+0.02}_{-0.01}$ &	0.44$^{+0.03}_{-0.02}$	\\
\\
 $\Gamma$	&	24.8$^{+1.14}_{-0.95}$		&	23.6$^{+1.76}_{-1.73}$	&	23.2$^{+2.4}_{-2.5}$	&	47$^{+8.6}_{-6.2}$
&	49.53$^{+9.5}_{-7.4}$ \\
\\
$U_{ph}$ & 2.7   & 3.55  &1.97 & 1.93 &4.77\\
\\
P$_{rad}$ &1.05&3.05&2.58&1.49&2.05\\
\\
P$_{jet}$ &2.62&2.48&3.85&4.1&20.2\\
\\

$\gamma_{min}$	&	250	&	303	&	200	&	200	&	200	\\
\\
\hline
$\chi^{2}$/d.o.f	&	39.5/13		&	24.7/13	&	29.8/13	&	36.11/13 &	23.99/13	\\
\hline\\
 \end{tabular}
\begin{tablenotes}
      \small
\item     $p, q$ = Low and high energy particle index\\ 
 $\gamma_b$ = Break Lorentz factor\\
 $B$ = Magnetic Field (in Gauss)\\
 $\Gamma$ = Bulk Lorentz factor\\ 
 $\gamma_{max}$ = 5$\times \rm 10^5$\\
 Equipartition, $\eta$ = 3\\
 Size of emission region,R = 2 $\times \rm 10^{16}$ cm\\
 Viewing angle, $\theta$ = 1 degree\\
 U$_{ph}$  = density of the external target photon
field (in $10^{-5} erg/cm^3$)\\
P$_{rad}$=Radiative power (in $10^{42}erg/s$)\\
P$_{jet}$=Jet power (in $10^{45}erg/s$)
 \end{tablenotes}
 \end{table*}
\section{Discussion}\label{sec:dis}

We studied the multi-wavelength behavior of OP 313 covering the period 60260-60400 in which it was detected in the VHE band with the Fermi-LAT and LST-1 and also exhibited multiple episodes of elevated activity. 
The patterns of \gm-ray flares were determined by fitting the exponential functions and measuring the flare rise and decay times. Following \citet[][]{2010ApJ...722..520A}, we considered the asymmetry parameter ($\zeta$) to define three classes of flares: (i) symmetric flares with $-0.3<\zeta <0.3$, (ii) moderately asymmetric flares with $-0.7<\zeta <-0.3$ or $0.3<\zeta <0.7$, and (iii) asymmetric flares with $-1<\zeta <-0.7$ or $0.7<\zeta <1$. Seven out of nine $\gamma$-ray flares appeared to be symmetric, whereas, two were moderately asymmetric in nature. The flare temporal characteristics is governed by the acceleration and energy loss time scales which is additionally modulated by the light travel time effects \citep{1998A&A...333..452K}. If the light travel time is much longer than the acceleration/loss timescale then one expects a symmetric flare with the profile governed by the geometry of the emission region \citep{1999MNRAS.306..551C}. On the other hand, if the acceleration and energy loss timescales are shorter/comparable to the light travel time, then its effects can cause asymmetric flares. For instance, a rapid acceleration against slow energy loss will result in fast-rise/slow-decay type flare profile; whereas, a slow-rise/fast-decay type flare profile suggests rapid energy loss against the acceleration. Since most of the flares observed are symmetric, it suggests that the light travel time governs the flare profile. Alternatively, a symmetric flare can also be obtained when the acceleration and the energy loss timescales are comparable. For the two moderately asymmetric flares, we find the decay time is shorter than the rise time and this suggests faster energy loss rate.

Symmetric flares advocate the light travel time to govern the flare temporal profile and hence the emission region size can be constrained from the shortest variability timescale ($t_{\rm var}$) and the Doppler factor ($\delta$) using light travel time argument as, 
 \begin{equation}
\label{eq:size}
R \leq \frac{c\,t_{\rm var}}{1+z}\, \delta
\end{equation}
The shortest \gm-ray flux doubling time was estimated to be 11.7 hours during MJD 60366-60367 (Table~\ref{tab:halving doubling time}). This period lies in the epoch of the brightest \gm-ray flare (P4) for which we estimated the Doppler factor to be 33.5, assuming $\delta\approx\Gamma$ (Table~\ref{tab:sed}). Inserting these parameters in the above equation, we get an upper limit to the emission region radius as \mbox{$2\times10^{16}$} cm. The same was used during the modeling of all the activity states.

Assuming a conical jet, the distance of the emission region from the central black hole can be estimated as \mbox{$D_{\rm blob} \sim 2 \,c\, t_{\rm var}\, \delta^2/(1+z)$} \citep{Abdo_2010}. Using the best-fit parameters for the epoch P4,  we get $D_{\rm blob}=1.4\times10^{18}$cm. Moreover, we can estimate the radii of the BLR and dusty torus respectively as
\begin{align}R_{\rm blr} = 10^{17}\,L_{\rm d,45}^{1/2}~\textrm{cm} \ \textrm{and} \ R_{\rm ir} = 2.5\,\times\, 10^{18}\,L_{\rm d,45}^{1/2}~\textrm{cm}
\end{align}
where, $ L_{\rm d,45}$ is the disc luminosity in 10$^{45}$ ergs s$^{-1}$ units \citep{Ghisellini_2009}. From the optical spectroscopic analysis, \citet{2021ApJS..253...46P} reported the disk luminosity of OP 313 to be 8.13$\times 10^{45}$ ergs s$^{-1}$. Inserting it in the above equations, we get $R_{\rm blr}=2.8\times10^{17}$ cm and $R_{\rm ir}=7.1\times10^{18}$ cm. On comparing $D_{\rm blob}$ with $R_{\rm blr}$ and $R_{\rm ir}$, we found the \gm-ray emitting region to be located outside the BLR but inside the dusty torus. This sets the torus photon field as the primary reservoir for the EC process. Indeed, the detection of photons with energy greater than 50 GeV photons from OP 313 also indicates the \gm-ray emission region to be located outside BLR to avoid \gm-rays getting absorbed by the intense BLR photon field via pair-production process \citep[e.g.,][]{2003APh....18..377D,2016ApJ...821..102B}.

The spectral behaviour of X-ray indicates that the dominent emission component is different at different epochs. We utilize this behaviour to calculate $\gamma_{min}$ and $\gamma_{max}$. Since the X-ray band fall on the rising part 
of the Compton spectral component during epochs Q, P2 and P4, these can be used to constrain $\gamma_{\rm min}$, while $\gamma_{\rm max}$
can be constrained from epoch P3 (falling section of the synchrotron spectral component). For epoch P1, the X-ray band fall on the transition regime
between synchrotron and Compton spectral components
and hence can be used to constrain both $\gamma_{\rm min}$ and $\gamma_{\rm max}$. Consistently, we use the observed photon frequencies to
estimate the $\gamma_{\rm min}$ and $\gamma_{\rm max}$ using \citep{2018RAA....18...35S}
\begin {equation}
 \label{gamin}
 \nu_{\rm syn,obs,max}=\frac{\delta}{1+z}\gamma_{\rm max}^2\nu_{L}
\end{equation}
and
\begin {equation}
 \label{gamax}
 \nu_{\rm ssc,obs,min}=\frac{\delta}{1+z}\gamma_{\rm min}^4\nu_{L}
\end{equation}
where, $\nu_L = eB/2\pi m_ec$ is the Larmor frequency.
From the best-fit parameters provided in Table~\ref{tab:sed}, the $\gamma_{min}$ values obtained for the epochs P1, P2, P4 and Q are $304$, $275$, $227$ and $245$, 
respectively. The calculated value of $\gamma_{max}$ obtained for epochs P1 and P3 are $0.9\times10^5$ and $2.6\times10^5$, respectively. The gyroradius
($R_{g}\simeq{\gamma m_{e}c^{2}}/{eB}$) corresponding to the maximum obtained $\gamma_{\rm max} \approx 2.6\times10^5$ is $1.31\times 10^9$ cm. This is much smaller than the size of the 
emission region estimated from the light travel time arguments and this supports that the electrons of such high energy can be constrained within the 
emission region.


The bulk kinetic power of the jet ($P_{\rm jet}$) can be estimated assuming the proton matter in the jet are cold (\citet[][]{2008MNRAS.385..283C}) and their number is 
equal to that of the non-thermal electrons 
\begin{equation}
\label{eq:jetpower}
P_{\rm jet}\approx\pi R^{2}\Gamma^{2}\beta_{\Gamma}U_{p}c
\end{equation}
where, $U_{\rm p}$ is the proton number density. From the best fit parameters we obtained the maximum and the minimum $P_{\rm jet}$ to be $6.09\times 10^{45} erg/s$ and $1.64\times 10^{45} erg/s$ respectively.  We also estimated the total radiated power through synchrotron and inverse Compton emission processes and are tabulated in Table \ref{tab:sed}.
We found the radiated power is much smaller compared to $P_{\rm jet}$ and hence most of the bulk energy is retained with in the jet after crossing the blazar emission zone. 





The broadband SED of the low-activity state (epoch Q) was well explained with the one-zone leptonic model used in this work. The optical-UV emission was reproduced with the synchrotron process. On the other hand, The X-ray and \gm-ray spectra were modeled with the SSC and EC-torus mechanisms, respectively. The obtained parameters, e.g., magnetic field and the bulk Lorentz factor, are similar to that derived for the low-activity state of another VHE detected FSRQ PKS 1441+25 \citep[see, e.g.,][]{2015ApJ...815L..23A}.


During the LST-1 detection epoch P1, we found OP 313 to be in an elevated activity state with respect to the low-activity period (Figure~\ref{combined_SED}). The synchrotron and EC-torus mechanisms explained the optical-UV and \gm-ray spectra, respectively. Interestingly, the X-ray spectrum revealed a concave upward shape. We reproduced the soft X-ray emission below $\sim$2 keV with the tail of the synchrotron process, while SSC process takes over at higher energies. This observation suggests a possible shift of the SED peaks to higher frequencies. Indeed, the break Lorentz factor was found to be larger than that determined for the low-activity state (Table~\ref{tab:sed}). This bluer-when-brighter behavior has been observed from several FSRQs \citep[e.g.,][]{2013MNRAS.432L..66G,2013MNRAS.435L..24T,2019A&A...627A.140A,2021ApJS..257...37P,2024JHEAp..42..115T}. In fact, similar spectral features have been identified during the VHE detection of FSRQs \citep[][]{2021A&A...647A.163M,2015ApJ...815L..22A}. Comparing with the low-activity state, the main changes in the SED parameters correspond to an increase in the magnetic field. The low energy slope of the broken power law electron energy distribution also decreased indicating a hardening of the spectral shapes.


During the activity period P2, which referred to the epoch of the Fermi-LAT VHE detection, OP 313 still had higher flux values compared to the low-activity state; however, it was in a relatively lower activity state with respect to the LST-1 detection epoch. Similar to other periods, the optical and \gm-ray spectra were well explained with the synchrotron and EC-torus processes, respectively. The X-ray spectrum was found to be soft during this period (Table~\ref{tab:swift}). From the modeling, it was found that the synchrotron mechanism contributed at soft X rays, whereas, SSC process can explain the hard X-ray emission. The \gm-ray emission had a hard spectral shape with the EC-torus marginally explaining the highest energy SED data point.



The activity period P3 was intriguing since OP 313 exhibited its brightest X-ray and optical flares during this period, while it was relatively moderately bright in the \gm-ray band (Figure~\ref{combined_SED}). The most interesting feature observed during this period was the detection of a steep falling X-ray spectrum, typically observed from high-synchrotron peaked BL Lac objects. The optical-UV and X-ray spectra were well reproduced with the synchrotron process. The EC-torus mechanism, on the other hand, explained the \gm-ray emission. Comparing the SED parameters, the break Lorentz factor was found to be higher than that at the low activity state, indicating a shift of the SED peaks to higher energies. Interestingly, this parameter was found to be similar to that obtained for the LST-1 detection epoch P1. The surprising fact is the vanishing of the concave shape observed during P1 and the overall synchrotron dominance in the X-ray band. These results highlight the complex radiative environment of OP 313. The slopes of the broken power law electron energy distribution became smaller, indicating the spectral hardening of the particle energy distribution.



The epoch P4 corresponded to the brightest \gm-ray flaring activity detected from OP 313 \citep[][]{2024ATel16497....1B}. From the modeling of the broadband SED, we found that the optical-UV emission is well explained with the synchrotron mechanism. The X-ray emission exhibited a flat rising spectral shape, typically observed in FSRQs, and was reproduced with the SSC process. The \gm-ray emission, on the other hand, was modeled with the EC-torus mechanism. However, the highest energy \gm-ray SED data point could not be explained, indicating a possible contribution from some other, more complex, radiative processes, e.g., hadronic component \citep[see, e.g.,][]{2024MNRAS.529.3503B}. Interestingly, the break Lorentz factor during this activity period was found to be greater than that obtained during the low-activity epoch, but smaller than those at other epochs, suggesting a small shift in the SED peaks to higher frequencies during the \gm-ray flare. This is also supported by similar X-ray shapes noticed during these two epochs. However, this is not an uncommon observation and several FSRQs have exhibited similar broadband spectral features during \gm-ray flaring episodes \citep[cf.][]{2015ApJ...803...15P,2015ApJ...807...79H}.



%
	%


\section{Summary}
This study, explores the multi-wavelength flaring activity of the distant flat spectrum radio quasar OP 313 (z=0.997) from November 2023 to March 2024. Using data from Fermi-LAT, Swift-XRT, and Swift-UVOT, we analyzed temporal and spectral characteristics to investigate jet emission processes and radiative mechanisms. The broadband spectral energy distribution (SED) revealed emissions driven by synchrotron radiation, synchrotron self-compton (SSC), and external compton (EC) processes, with torus photons dominating the EC mechanism. The \gm-ray emitting region was deduced to be outside the broad-line region (BLR) but within the dusty torus. Significant X-ray spectral variability highlighted a dynamic radiative environment, and analysis of \gm-ray light curves identified both symmetric and asymmetric flares, indicating variability induced by light travel time effects and particle acceleration. Further, the broadband spectral modeling showed that the bulk of the jet's energy remains intact beyond the emission zone, emphasizing efficient energy retention. These findings advance the understanding of particle acceleration, photon field interactions, and jet dynamics in flat spectrum radio quasars.


\section*{Acknowledgements}
We gratefully acknowledge the use of data from the Fermi Science Support Center (FSSC) and Swift data from the High Energy Astrophysics Science Archive Research Center (HEASARC) at NASA, Goddard Space Flight Center. NMPN sincerely appreciates the research facilities provided by UGC-SAP and FIST 2 (SR/FIST/PS1-159/2010) (DST, Government of India) in the Department of Physics, University of Calicut. We also extend our gratitude to DST-FIST for providing research facilities at Farook College (Autonomous), Calicut.
\section*{Data Availability}
The data utilized in this study are publicly accessible and were obtained from the archives available at \url{https://heasarc.gsfc.nasa.gov/} and \url{https://Fermi.gsfc.nasa.gov/}.




\bibliographystyle{elsarticle-num-names} 
\bibliography{cas-refs}



\end{document}